\documentclass{ws-ijmpcs}

\begin{document}

\markboth{I.O. Cherednikov, T. Mertens, P. Taels, F.F. Van der Veken}
{Evolution of TDDs at large-$x_B$ and geometry of the loop space}

%
\catchline{}{}{}{}{}
%

\title{EVOLUTION OF TRANSVERSE-DISTANCE DEPENDENT PARTON DENSITIES AT LARGE-$X_B$ AND GEOMETRY OF THE LOOP SPACE \footnote{Presented at the QCD Evolution Workshop, JLab, May 6th-10th, 2013.}
}

\author{IGOR O. CHEREDNIKOV\footnote{
Speaker}}

\address{Departement Fysica, Universiteit Antwerpen,
Antwerp, B-2020 Belgium\\ and \\
Bogoliubov Laboratory of Theoretical Physics, JINR,
RU-141980 Dubna, Russia\\
igor.cherednikov@uantwerpen.be}

\author{TOM MERTENS}

\address{Departement Fysica, Universiteit Antwerpen,
Antwerp, B-2020 Belgium\\
tom.mertens@uantwerpen.be}

\author{PIETER TAELS}

\address{Departement Fysica, Universiteit Antwerpen,
Antwerp, B-2020 Belgium\\
pieter.taels@uantwerpen.be}

\author{FREDERIK F. VAN DER VEKEN}

\address{Departement Fysica, Universiteit Antwerpen,
Antwerp, B-2020 Belgium\\
frederik.vanderveken@uantwerpen.be}

\maketitle

\begin{history}
\received{\today}
\end{history}

\begin{abstract}

We discuss possible applications of the equations of motion in the generalized Wilson loop space to the phenomenology of the three-dimensional parton distribution functions in the large-$x_B$ approximation. This regime is relevant for future experimental programs to be launched at the (approved) Jefferson Lab $12$ GeV upgrade and the (planned) Electron-Ion Collider. We show that the geometrical evolution of the Wilson loops corresponds to the combined rapidity and renormalization-group equation of the transverse-distance dependent parton densities in the large-$x_B$ factorization scheme.

\end{abstract}

\ccode{PACS numbers: 11.25.Hf, 123.1K}

\section{Introduction: Shape variations in the loop space and renormalization-group evolution}	

Generalized loop space, associated to some base manifold, consists of the distributional version\footnote{In the sense that one needs to integrate over the integration contour or path ($\{\Gamma_i\}$) in the base manifold.} of closed paths, with an equivalence relation introduced by the holonomy of the gauge connection at the base point of the closed paths, and where taking the trace provides gauge invariance, see, e.g., [\refcite{General_LS}] and references therein.
This holonomy is also referred to as the  Wilson loop, Eq.  (\ref{eq:WL_definition}), where the gauge fields ${\cal A}_\mu$ belong to a certain representation of the non-Abelian gauge group  $SU(N_c)$ and where the closed paths $\{\Gamma_i\}$ are elements of generalized loop space. Taking vacuum expectation values over a set of Wilson loops leads to Wilson loop variables, Eq. (\ref{eq:wl_def}), which are an alternate way of representing a gauge theory [\refcite{General_LS}] under certain constraints known as the Mandelstam constraints, Eq. (\ref{eq : mandelstam constraints}):

\begin{eqnarray}
 & & {\cal W}_n [\Gamma_1, ... \Gamma_n]
  =
\Big \langle 0 \Big| {\cal T} \frac{1}{N_c} {\rm Tr}\ \Phi (\Gamma_1)\cdot \cdot \cdot \frac{1}{N_c}{\rm Tr}\ \Phi (\Gamma_n)  \Big| 0 \Big\rangle \ , \label{eq:wl_def} \\
 & & \Phi (\Gamma_i)
   =
   {\cal P} \ \exp\left[ig \oint_{\Gamma_i} \ dz^\mu {\cal A}_{\mu} (z) \right] \ .
   \label{eq:WL_definition}
\end{eqnarray}
These variables are now {\it functionals on loops}, and exhibit non-trivial behaviour in the vicinity of path discontinuities, cusps or self-intersections. Moreover, the renormalization and conformal properties of  Wilson loops possessing light-like segments (or lying completely on the light-cone) are known to be more intricate than their off-light-cone counterparts. Therefore, the study of the
geometrical and dynamical properties of loop space, which can include, in general, cusped light-like Wilson exponentials, will provide us with fundamental information on the renormalization group behaviour and evolution of the various gauge-invariant quantum correlation functions [\refcite{Loop_Space,WL_RG}].

Generalized loop space can be shown to be a topological group that has an (infinite dimensional) Lie algebra
associated with it, allowing the definition of different differential operators such as the path- and area-derivative, Eqs. (\ref{eq:area_derivative},\ref{eq:path_derivative}):
\begin{align}
 	\frac{\delta}{\delta \sigma_{\mu\nu} (x)} \ \Phi (\Gamma)
 	&\equiv
 	\lim_{|\delta \sigma_{\mu\nu} (x)| \to 0} \ \frac{ \Phi (\Gamma\delta \Gamma) - \Phi (\Gamma) } {|\delta \sigma_{\mu\nu} (x)|} \ ,
	\label{eq:area_derivative}\\
	 \partial_\mu  \Phi(\Gamma)
 	&=
 	\lim_{|\delta x_{\mu}| \to 0} \frac{\Phi(\delta x_\mu^{-1}\Gamma\delta x_\mu) - \Phi(\Gamma)}{|\delta x_{\mu}|} 		\ ,
 	\label{eq:path_derivative}
\end{align}
used by Makeenko and Migdal to derive their famous non-perturbative equations [\refcite{MM_WL,WL_Renorm}]:
\begin{equation}
 	\partial_x^\nu \ \frac{\delta}{\delta \sigma_{\mu\nu} (x)} \ W_1[\Gamma]
 	=
	 N_c g^2 \ \oint_{\Gamma} \ dz^\mu \ \delta^{(4)} (x - z) W_2[\Gamma_{xz} \Gamma_{zx}] \ ,
 \label{eq:MM_general}
\end{equation}
supplemented with the Mandelstam constraints
\begin{equation}
	\sum a_i { {\cal W}_{n_i} [\Gamma_{1} ... \Gamma_{n_i}] } = 0 \ .
	\label{eq : mandelstam constraints}
\end{equation}

However, the possibility to apply the MM equations directly in the form (\ref{eq:MM_general}) suggest the use of sufficiently smooth paths off-the-light-cone, which excludes, e.g., loops with cusps and loops (partially) lying on the light-cone. Such paths occur, e.g., in the investigation of the duality between the $n-$gluon scattering amplitudes and $n-$polygonal Wilson loops in ${\cal N} = 4$ super-Yang-Mills theory, Refs. [\refcite{WL_CFT}], or in the configurations arising in the soft parts of $3D$-parton densities (see below).
In recent work, Refs. [\refcite{ChMVdV_2012}], we developed an approach which allows one to apply the Schwinger method [\refcite{Schwinger51}], as used in the derivation of the Makeenko-Migdal equations,  to certain classes of cusped loops (partially) lying on the light-like rays. To this end, we defined  a new differential operator, Eq. (\ref{eq:area_log}), which might be related to the Fr{\'e}chet derivative [\refcite{General_LS}], where loop variations are generated by an infinite number of area-derivative-like variations.
This approach led us to propose an evolution equation valid for the planar light-like Wilson rectangular loops:
\begin{figure}[ht]
 $$\includegraphics[angle=90,width=0.6\textwidth]{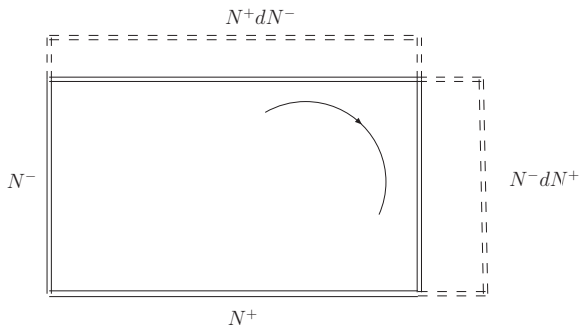}$$
   \caption{Shape variations allowed for a light-cone Wilson rectangle: we consider only those variations of the integration path which conserve the angles between the sides.}
\end{figure}

\begin{equation}
 \mu \frac{d }{d \mu } \ {\left(  \frac{d}{d \ln \sigma} \ \ln \ {\cal W} [\Gamma] \right) }
 =
 - \sum { \Gamma_{\rm cusp} } \ ,
 \label{eq:mod_schwinger}
\end{equation}
where the area differentials are defined in the transverse $\bm z_\perp = 0$:
\begin{equation}
{ d \sigma^{+-} }
=
{ N^+ d N^- } \  , \
 { d \sigma^{-+} }
=
- { N^- d N^+ }  \ ,
\label{eq:delta_area}
\end{equation}
and the only allowed shape variations are presented in Fig. 1. The  area logarithmic derivation operator then reads
\begin{eqnarray}
  \frac{d}{d \ln \sigma}
  \equiv
  \sigma^{\mu\nu} \frac{d}{d \sigma^{\mu\nu}}
  =
  \sigma^{+-} \frac{d}{d \sigma^{+-}}
  +
  \sigma^{-+} \frac{d}{d \sigma^{-+}} \ .
  \label{eq:area_log}
\end{eqnarray}
The r.h.s. of Eq. (\ref{eq:mod_schwinger}) is defined by the sum of the light-cone cusp anomalous dimensions, the total amount of which is given by the number of the cusp-like discontinuities in the slope of the integration path. The  sum of the cusp anomalous dimensions [\refcite{CAD_universal}] in the r.h.s. of Eq. (\ref{eq:mod_schwinger}) can be interpreted as a fundamental ingredient of an effective quantum action for the Wilson loops with simple obstructions.
Hence, we reveal a connection between the geometrical properties (given in terms of the area/shape infinitesimal variations and corresponding differential equations) of the generalized loop space, and the renormalization-group behaviour of the Wilson polygons with conserved angles between the light-like straight lines (that is to say, classical conformal invariance is assumed). In other words, the dynamics in loop space is governed by the discontinuities of the path derivatives. These obstructions play the role of the {sources} within the Schwinger field-theoretical picture. We have shown, therefore, that the Schwinger quantum dynamical principle [\refcite{Schwinger51}] is helpful in the investigation of certain classes of  elements of the loop space, namely, the planar cusped Wilson polygons. It is worth noting that Eq. (\ref{eq:mod_schwinger}) suggests, in fact, a duality relation between the rapidity evolution of certain correlation functions and the equations of motion in the generalised loop space. Rapidities associated with the light-like vectors $N^\pm$ are, of course, infinite, and are given by
\begin{equation}
 Y^\pm
 =
 \frac{1}{2} \ \ln \frac{(N^\pm)^+}{(N^\pm)^-}
 =
\lim_{\eta^\pm \to 0}  \pm \frac{1}{2} \ \ln \frac{N^+N^-}{\eta^\pm} \ ,
\label{eq:rapidity}
\end{equation}
where $\eta^\pm$ is a cutoff and where we take into account the fact that plus- and minus- components of a vector $a_\mu$ are defined as the scalar products $a^\pm = (a \cdot N^\mp)$.
Eq. (\ref{eq:rapidity}) suggests, clearly, that
\begin{equation}
\frac{d }{d \ln \sigma} \sim \frac{d }{d Y} \ ,
\end{equation}
that is, the rapidity evolution equation of a certain correlation function can be set dual to the area variation law of a properly chosen class of elements of the generalized loop space.
In particular, we will argue below that Eq. (\ref{eq:mod_schwinger}) can be used to study the evolution of the $3D$-parton distribution functions.
To this end, we will focus on the behaviour of parton densities in the large Bjorken-$x_B$ approximation.

\section{Large-$x_B$ regime at modern experimental facilities}
Describing the three-dimensional structure of the nucleon is one of the hot topics of a number of current and planned experimental and theoretical projects (see [\refcite{INT,JLab12GeV,TMD_Pheno}] and references therein).
In the infinite-momentum frame (IMF), the nucleon can be treated as a complicated quantum many-body system of point-like constituents (partons). These partons can be probed in high-energy scattering processes with a spatial resolution equal to the inverse momentum transfer $Q^{-1}$. The energy transfer is then related to the Bjorken variable $x_B$, which defines as well the fraction of the nucleon's longitudinal momentum carried by the struck parton.
The region of $0.1 < x_B <1$ features the predominance of the three valence quarks inside the nucleon, while the effects of sea quarks become significant in the region of $0.01 < x_B < 0.1$, and gluons are known to dominate at even smaller $x_B$.

Past studies of fully inclusive deep inelastic scattering (DIS) heavily relayed on the IMF framework, using factorization to separate out IR and collinear singularities, the latter being collected from the hard (perturbatively calculable) part into the parton distribution functions (PDFs). In this case, it suffices to treat the struck parton as collinear with the parent nucleon. Deep inelastic electron-nucleon scattering experiments deliver, therefore, the information about the one-dimensional (longitudinal) structure of the nucleon.  In semi-inclusive processes (such as Drell-Yan, SIDIS, etc.), however, it is necessary to additionally incorporate the intrinsic transversal degrees of freedom of the struck partons, and in specific processes one also has to add possible spin-correlations to the PDFs. This information is delivered by the transverse momentum distributions (TMDs), providing a full three-dimensional picture of the nucleon structure.

A promising setup to investigate the behaviour of $3D$ parton densities at large $x_B$ is the new energy upgrade from 6 to 12 GeV of CEBAF at Jefferson Lab, which is being effectuated at the moment [\refcite{JLab12GeV}].  CEBAF is a fixed-target experiment, so one has $x_B=Q^2 / (2M\nu)$, implying that this upgrade will allow us to probe the region $x_B$ from $\sim 0.1$  up to very close to 1, in other words one will probe mainly valence quarks. This is ideal to study the evolution of density functions in the so far unexplored large-$x_B$ region. On the other hand, smaller $x_B$ (in the same region of values of $Q^2$) can be reached by the planned EIC, which will target sea quarks and gluons as well [\refcite{INT}]. The combined kinematic range of both experiments will be of the order to $10^{-3}  < x_B < 1$ and $2$ GeV$^2 < Q^2 < 100$ GeV$^2$. This coverage is unique in its large-$x_B$ component, which will allow us to make precision tests of TMD factorization and evolution. For the extended original discussion and figures, see [\refcite{JLab12GeV}] and Refs. therein.

\section{Large-$x_B$ factorization and evolution of transverse-distance dependent parton densities}

Three-dimensional parton densities in the momentum representation (TMDs) are intensively studied within different frameworks, Refs. [\refcite{TMD_CSS,TMD_origin,CS_all,New_TMD_Col,SCET_TMD,TMD_Pheno}]. It is beyond the scope of this paper to compare advantages and drawbacks of those approaches. Instead, we will just take it for granted that some reasonable TMD-factorization scheme can be formulated, and that an appropriate operator definition of the TMD quark distribution function exists that has the same quantum numbers as the correlation function below, Eq. (\ref{eq:TDD_general}).
In the present discussion we will be particularly concerned with the $3D$-correlation functions in the large-$x_B$ limit: this regime is apparently easier to analyze within a simple factorization scheme and perfectly fits the Jefferson Lab 12 GeV kinematics. Furthermore, we will argue that the large-$x_B$ approximation is an ideal natural laboratory for the study of applications of the generalized loop space formalism in hadronic and nuclear physics.

To this end, let us consider the following {\it transverse-distance dependent} (TDD) correlation function
\begin{eqnarray}
& & {\cal F} \left(x, {\bm b}_\perp; P^+, n^-, \mu^2 \right)
=  \int\! d^2 k_\perp \  {\rm e}^{-ik_\perp \cdot b_\perp} {\cal F} \left(x, {\bm k}_\perp; p^+, n^-, \mu^2 \right) =  
\label{eq:TDD_general} \nonumber \\
& &   \int\! \frac{d z^-}{2\pi} \left\langle
              P \ | \bar \psi (z^-,  \bm{b}_\perp)
              {\cal W}_{n^-}^\dagger[z^-,  \bm{b}_\perp;
   \infty^-,  \bm{b}_\perp] {\cal W}_{\bm l}^\dagger[\infty^-,  {\bm b}_\perp;
   \infty^-,  {\infty}_\perp] \right.   \\
   && \left.
\times
   \gamma^+ {\cal W}_{\bm l}[\infty^-,  {\infty}_\perp;
   \infty^-, \bm{0}_\perp]
   {\cal W}_{n^-}[\infty^-, \bm{0}_\perp; 0^-,\bm{0}_\perp]
   \psi (0^-,\bm{0}_\perp) | \ P
   \right\rangle \nonumber \ ,
\end{eqnarray}
which is supposed to deliver the information about the quark distribution in the longitudinal one-dimensional momentum space and in the two-dimensional impact-parameter coordinate space.
Generic semi-infinite Wilson lines evaluated along a certain four-vector $w_\mu$ are defined as
\begin{equation}
\label{eq:SIWL}
{\cal W}_w[\infty ; z]
\equiv {}
  {\cal P} \exp \left[
                      - i g \int_0^\infty d\tau \ w_{\mu} \
                      {\cal A}^{\mu} (z + w \tau)
                \right] \ ,
\end{equation}
where, in the cases under consideration, the vector $w_\mu$ can be either longitudinal $w_\mu = (w_L, \bm 0_\perp)$, or transverse $w_\mu = (0_L, {\bm l}_\perp)$.
The TDD arises, by definition, as a result of the partial Fourier transform of the standard {\it transverse-momentum dependent} correlator ${\cal F} \left(x, {\bm k}_\perp; P^+, n^-, \mu^2 \right)$.
The factorization and evolution of the gauge-invariant collinear PDFs in the large-$x_B$ regime has been studied in Ref. [\refcite{LargeX_KM}]. We propose to generalize this approach to the $3D$-PDF, Eq. (\ref{eq:TDD_general}).

The large-$x_B$ regime suggests the following assumptions (for a detailed discussion, see Ref.[\refcite{LargeX_KM,CMTV_LargeX}]):

\begin{itemize}

\item In the IMF, the struck quark acquires ``almost all'' the momentum of the nucleon, that is: $k_\mu \approx P_\mu$. Provided that the transverse component of the nucleon momentum is equal to zero, the transverse momentum of the quark $\bm k_\perp$ is gained by the gluon interactions;

\item A very fast moving quark with momentum $k_\mu$ can be considered as a classical particle with a (dimensionless) velocity parallel to the nucleon momentum $P$, so that the quark fields are replaced by
$$
\psi (0)
=
{\cal W}_P [\infty; 0] \ \Psi_{\rm in-jet} (0)
\  , \ \bar \psi (z^-, \bm z_\perp)
=
\bar \Psi_{\rm in-jet}  (z) \ {\cal W}_P^\dag [z ; \infty] \ ,
$$
where the fields $\bar \Psi_{\rm in-jet}, \Psi_{\rm in-jet}$ represent the incoming-collinear jets in the initial and final states [\refcite{Eikonal}];

\item Provided that ``almost all'' momentum of the nucleon is carried by the struck quark, real radiation can only be soft: $q_\mu \sim (1- x) P_\mu$;

\item Virtual gluons can be soft or collinear, collinear gluons can only be virtual, quark radiation is suppressed in the leading-twist;

\item Rapidity singularities stem only from the soft contributions: they are known to occur in the soft region, at small gluon momentum $q^+ \to 0$. In other words, rapidity divergences are known to originate from the minus-infinite rapidity region, where gluons travel along the direction of the outgoing jet, not incoming-collinear;

\item Real contributions are UV-finite (in contrast to the integrated PDFs), but can contain rapidity singularities and a non-trivial $x_B$- and $\bm b_\perp$-dependence.

\end{itemize}

These assumptions imply the following large-$x_B$ factorization formula:
\begin{equation}
{\cal F} \left(x, {\bm b}_\perp; P^+, n^-, \mu^2 \right)
=
{\cal H} (\mu, P^2) \times {\Phi} (x, {\bm b}_\perp; P^+, n^-, \mu^2 ) \ ,
\label{eq:LargeX_factor} 
\end{equation}
where the contribution of incoming-collinear partons is summed up into the $x_B$-independent function, while
the soft function $\Phi$ is given by\footnote{For the sake of simplicity, we work in covariant gauges, so that the transverse Wilson lines at infinity can be ignored.}
\begin{eqnarray}
& & {\Phi} (x, {\bm b}_\perp; P^+, n^-, \mu^2 )
= P^+ \int\!dx \ {\rm e}^{-i (1-x) P^+ z^-} \cdot \nonumber \\
& &  \times \langle 0 | \ {\cal W}_P^\dag [z ; - \infty] {\cal W}_{n^-}^\dag[z; \infty] {\cal W}_{n^-} [\infty ; 0] {\cal W}_P [0; \infty] \  | 0 \rangle \ ,
\label{eq:soft_LargeX}
\end{eqnarray}
with two kinds of Wilson lines: incoming-collinear (non-light-like, $P^2 \neq 0$) ${\cal W}_P$, and outgoing-collinear ($(n^-)^2 = 0 $), ${\cal W}_{n^-}$.

The rapidity and renormalzation-group evolution equations are:
\begin{eqnarray}
  & & \mu \frac{d}{d\mu} \ln {\cal F} \left(x, {\bm b}_\perp; P^+, n^-, \mu^2 \right)
  = \mu \frac{d}{d\mu} \ln {\cal H} (\mu^2) + \mu \frac{d}{d\mu} \ln {\Phi} (x, {\bm b}_\perp; P^+, \mu^2)
  \label{eq:TDD_evolution_1}
  \ , \\
  & & P^+ \frac{d}{d P^+} \ln {\cal F} \left(x, {\bm b}_\perp; P^+, n^-, \mu^2 \right)
  = P^+ \frac{d}{d P^+} \ln {\Phi} (x, {\bm b}_\perp; P^+, \mu^2)  \ .
  \label{eq:TDD_evolution_2}
  \end{eqnarray}
Note that the rapidity is introduced via $\ln P^+$ with proper regularization [\refcite{Li_WL}]. The r.h.s. of Eq. (\ref{eq:TDD_evolution_1}) is, in fact, $\bm{b}_\perp$-independent and contains only a single-log dependence on the rapidity  [\refcite{CS_all}]. Therefore, the r.h.s. of Eq. (\ref{eq:TDD_evolution_2}) corresponds to the Collins-Soper-Sterman rapidity-independent kernel ${\cal K}_{\rm CSS}$. At this point, we are in a position to make use of the evolution equation (\ref{eq:mod_schwinger}). To this end, we emphasize that the soft function ${\cal F}$ is a Fourier transform of an element of the loop space: the Wilson loop evaluated along the path, defined in Eq. (\ref{eq:soft_LargeX}). This fact enables us to consider the shape variations of this path, which are generated by the infinitesimal variations of the rapidity variable $\ln P^+$. The corresponding differential operator reads
$$
\frac{d}{d \ln \sigma} \sim P^+ \frac{d}{d P^+} \ ,
$$
given that $d P^+ = (dP \cdot n^-)$. Therefore:
\begin{equation}
 \mu \frac{d}{d\mu} \ \left(P^+ \frac{d}{d P^+} \ln {\cal F} \right)
 = \mu \frac{d}{d\mu} \ \left( P^+ \frac{d}{d P^+} \ln {\Phi} \right)
 =
 - \sum_{\rm TDD} \Gamma_{\rm cusp} (\alpha_s )
 =
 \mu \frac{d}{d\mu}
  {\cal K}_{\rm CSS}(\alpha_s ) \ .
  \label{eq:CSS}
\end{equation}
Equations (\ref{eq:TDD_evolution_1}-\ref{eq:CSS}) can be straightforwardly integrated to give a complete evolution of the TDD (\ref{eq:TDD_general}) in the large-$x_B$ region [\refcite{CMTV_LargeX}], which can be directly applied to the JLab 12 GeV phenomenology (see also [\refcite{Accardi_2013}] and Refs. therein).

\section{Conclusions}
We have shown that the geometrical properties of the generalized loop space can be used to analyze the combined rapidity and RG-evolution of the transverse-distance dependent parton densities in the large-$x_B$ approximation. This regime allows one to factorize the soft contributions responsible for the non-trivial $x_B$- and $\bm b_\perp$-dependence in the form of  vacuum averages of the system of Wilson lines lying on- and off- the light-cone. It is crucial that the rapidity divergences get factorized into the soft part as well, while incoming-collinear jets contribute to the renormalization-group evolution via the corresponding anomalous dimensions. Being an element of the generalized loop space, the soft function $\Phi$ obeys the equations of motion, which describe its behaviour under the shape variations of the specially chosen form. The latter can be connected with the rapidity differentials in the energy-momentum space.
Although the differential equations, Eqs. (\ref{eq:TDD_evolution_1}-\ref{eq:CSS}), coincide with the ones obtained in other approaches [\refcite{TMD_CSS,New_TMD_Col,SCET_TMD}], their solution seems to be easier due to the simpler structure of the integration constants and the more straightforward relation to the integrated case. The solution and applications will be reported separately [\refcite{CMTV_LargeX}].

\section*{Acknowledgments} We thank V. Guzey for interest in our work, and for numerous useful comments.


\end{document}